\documentstyle[floats,aps]{revtex}
\input{psfig}
\begin{document}
\def \beq{\begin{equation}}
\def \eeq{\end{equation}}
\def \beqarr{\begin{eqnarray}}
\def \eeqarr{\end{eqnarray}}
\def \be{\begin{equation}}
\def \ee{\end{equation}}
\def \bea{\begin{eqnarray}}
\def \eea{\end{eqnarray}}
\def \ta{{\tilde\alpha}}
\def \tg{{\tilde g}}     
\twocolumn[\hsize\textwidth\columnwidth\hsize\csname @twocolumnfalse\endcsname
\title{Transport Through Quantum Melts}
\author{Efrat Shimshoni$^{1}$,  Assa Auerbach$^2$  and Aharon Kapitulnik$^3$}
\address{$^1$ Department of Mathematics-Physics, Oranim--Haifa University,
Tivon 36006, Israel.}
\address{$^2$ Department of Physics, The Technion, Haifa 32000, Israel.}
\address{$^3$ Department of Applied Physics, Stanford University,
Stanford, CA 94305}
\date{\today}
\maketitle
\begin{abstract}
We discuss superconductor to insulator
and quantum Hall transitions which are first  order in the clean limit. 
Disorder creates
a nearly percolating network of the  minority phase. Electrical transport
 is dominated by tunneling or activation through the saddle point junctions, 
whose typical resistance is calculated as a function of magnetic field.  
In the Boltzmann regime, this approach yields
resistivity  laws which agree with recent experiments in both classes of 
systems. We discuss the origin of dissipation at zero temperature.
\end{abstract}
\vskip2pc]
\narrowtext
Two--dimensional (2D) electron systems subject to disorder potentials and
external fields exhibit a rich set of quantum phase
transitions, indicated by drammatic changes in their transport
properties at low temperatures.  Here we concentrate 
on two  prominent classes (i)  Superconductor to Insulator (S--I)
transitions \cite{FisherLee,SITrev,YK,Krav} observed in a variety of 
superconducting films and in Josephson arrays, and typically tuned by either 
disorder or magnetic field. (ii) Analogous
transitions in the Quantum Hall  (QH) regime: the QH to insulator (QH--I) 
transition, and transitions between different 
QH plateaux\cite{KLZ,phas-tran,Shahar1}. 

The longitudinal
sheet resistivity $\rho_{xx}$ in these systems is
a continuous function of $T$, $B$ and $n$, the  temperature, magnetic field  
and carrier density respectively.    
A sharp change in $\lim_{T\to 0}\rho_{xx}$ as a function of $B$ 
has been  interpreted as a quantum phase transition, between
localized bosons and localized vortices\cite{FisherLee,KLZ}. 

Recent experiments, however, found a remarkably simple {\em non critical} 
behavior of the resistivity which seems to hold  in a sizeable portion of the 
phase diagram.\newline
(i) On both sides of the QH--I transition\cite{Shahar2}
\beq
\rho_{xx}={h\over e^2} 
\exp\left\{ \frac{-(\nu-\nu_{c})}{\alpha T + \beta} \right\} ,
\label{rlaw}
\eeq
where $\nu=n\phi_0/B$ (with $\phi_0$ the flux quantum) is
the average Landau level filling factor\cite{nuprime}, and $\nu_{c}$ is
its value at the critical point. 
$\alpha$  and $\beta$ are sample specific parameters.

(ii)  Near the field--tuned  S--I transition \cite{YK}:
\beq
\rho_{xx}={h\over 4e^2}\times
\cases{  \exp\left( \frac{B-B_{cr}}{ {\bar\alpha} T} \right)& large $T$\cr
 \exp\left( \frac{B-B_{cr}}{ {\bar\beta}  } \right) & $T \to 0$} 
\label{rlawB}
\eeq
where $B_{cr}$,   ${\bar \alpha}$ and ${\bar \beta}$ are constants.
Note that both Eq. (\ref{rlaw}) and (\ref{rlawB})
indicate finite  dissipation  at $T=0$ at all magnetic fields.

It is the purpose of this Letter to provide an interpretation of these
resistivity laws using Boltzmann transport theory of a binary composite
of two phases: conducting ($C$) and an insulating ($I$). 
The primary underlying assumption of this approach, is that  
{\em without disorder, the thermodynamic transition at $T=0$ is first order}.
Mathematically,  a crossing of
two ground state energy surfaces, $E_C(B,\mu)$ and $E_I(B,\mu)$ is assumed.
Here  
$\mu$ is the chemical potential of the charge carriers.  
The surface cross at a critical line 
$\mu_{cr}(B_{cr})$. 
Associated with the two phases are finite size correlation lengths  
$\xi_i, i=C,I$. These provide the  
lower limit to the linear size of an ordered  domain.

A smooth random potential $V(x,y)$, $\langle V\rangle=0$,  
with fluctuation lengthscale $l_V>\xi_i$ can be incorporated as a local 
shift in the chemical potential, such that the  local energy density is 
$e_i(B,\mu-V(x,y))$.  A large $\langle V^2 \rangle$ breaks  the system into 
domains which are approximately bounded by equipotential contours 
$V(x_\mu,y_\mu)=\mu-\mu_{cr}(B)$.  In QH 
systems, detailed calculations indicate phase separation \cite{Herb}
and domain sizes have been estimated
\cite{Chk}.  

The first order ``quantum melting'' assumption is  supported by 
theoretical arguments and some direct experimental evidence.

The theoretical models describing this type of systems exhibit a
competition between superconductivity and charge density correlations, as well
captured by their mapping to
an anisotropic $XXZ$  pseudospin model on a lattice. Sizeable portions
of parameter space for bipartite\cite{AA} 
and frustrated lattices\cite{MAA}, yield  first order 
transitions between solid and  superfluid phases\cite{AA,MAA,LG}.
Even when the classical transition is of second order, quantum corrections
can make it first order\cite{HLM}. 
A similar result was found for the Chern-Simons field theory of
the QH problem\cite{PZ}.  

An experimental evidence for a quantum melt scenario is provided by
photoluminescence (PL) data in QH systems \cite{Kuk}, which show
two distinct modes of relaxation within the sample. These
are interpreted in terms of sample inhomogeneity due to
binary phase separation. 

The assumption of a binary composite structure has been also used to explain
non universal critical conductivity in QH transitions\cite{Harvard}, and the 
quantization of the Hall resistivity 
at the QH-I transition\cite{ruzin} and in the QH insulator phase\cite{SA}.

The random potential 
eliminates the first order thermodynamic transition and  produces  
a second order transition of the transport coefficients which is of a 
percolative nature \cite{Berg,KSCC}. This accounts for many universal features
observed in different transitions. It can also help explain at least 
qualitatively, a duality relation observed when $C$ and $I$
phases are interchanged across the transition
\cite{Shahar1,Shahar2,SITdual,cldual}. 

The primary contribution to the resistivity
comes from  saddle points of the potential near $V(x,y)=\mu_{cr}$.
Here we concentrate on the Boltzmann regime, where it is implicitly assumed 
that incoherent scattering occurs within a single domain size. 
This requires sufficient zero temperature dissipation, a point
we shall return to in the end. 
Boltzmann theory uses the current density and electric field as classical 
variables which depend locally on each other.   
For a finite width distribution of  junction resistances in a 
two dimensional array, the total resistance is 
given by the resistance of  the typical junction \cite{SA}.  
 
A saddle point junction has two domains separated by minimal distance $d$.
The Ohmic response depends on
the transition rate ${\cal T}$ of the relevant quasiparticles which pass 
through the junction. 
\be
 {\cal T} \sim \cases{  \exp\left(-{V'' d^2\over 8 T}\right) & large $T$\cr
 \exp\left(-  { S'' d^2\over \hbar} \right)  & $T \to 0$} 
\label{Tdtun}
\ee
where $V''$ cand $S''$ are the curvatures of the potential barrier and 
tunneling action respectively. 

The resistivity of a single junction
is given by:
\beq
R_{xx}\sim{h\over Q^2}{1-{\cal T}\over {\cal T}}\; .
\label{rland}
\eeq
In the  insulating  side of the percolative transition,   
quasiparticles which flow between superconducting domains are charge $Q=2e$ 
Cooper pairs (bosons),
and for QH domains, they are electrons ($Q=e$) in the lowest Landau level.  

In the conducting side, the  quasiparticles  are
of vortices or edge quasiparticles which tunnel with rate ${\bar {\cal T}}$ 
between edges of a narrow superconducting
or QH liquid channel respectively. Since a current of vortices produces a 
longitudinal voltage drop,
the channel's resistance is given by the inverse expression to (\ref{rland})
\beq
R_{xx}\sim{h\over Q^2}    { {\bar {\cal T}}\over  1-{\bar {\cal T}} }\; .
\label{rlandv}
\eeq
A recent calculation\cite{Assa} of
the quasiparticle tunneling rate across a  quantum Hall  strip has found
$  S''  =   Q {\pi\over 4 l^2}d^2 $ (where $l^2=\hbar c/(eB) $)
for quasiparticles of charge $Qe$ for the QH liquid. 
For vortex tunneling through a superconductor there
are two limits which depend on the vortex core dissipation
\cite{RMP}: 
When dissipation due to the normal core is negligible, 
vortices obey "Hall" dynamics and $S'' \approx  \pi^2\rho_s/ 2  $
where $\rho_s$ is the superfluid density. 
In the opposite, viscous dynamics  limit,   
$S''\approx  \eta$ where the viscosity of the normal core is
given by Bardeen and Stephen\cite{BS} as 
\be
\eta=\hbar^2 /(2\pi \xi^2 e^2 \rho_n)
\label{eta}
\ee
where $\xi$ is the vortex core size, 
and $\rho_n$ is the normal state resistance measured
above the bulk superconducting transition temperature.

In order to compare theory to experimental results (\ref{rlaw},\ref{rlawB}),
the typical junction width  $d$  as a function  of external magnetic field
is required. These can be derived by  geometrical arguments.
We start with the QH case.

\begin{figure}[htb]
\centerline{\psfig{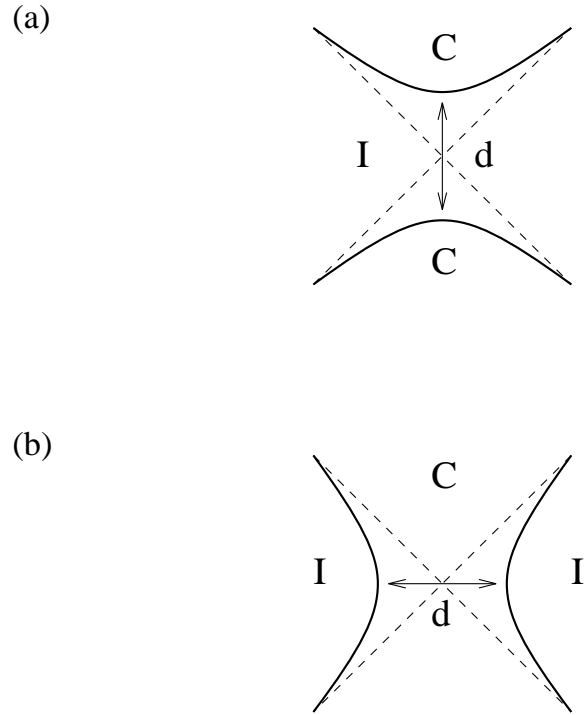}}
\vspace{0.5in}
\caption{
\label{fig:fig1}
A typical junction in a C--I mixture (a) in the insulating phase, and (b) in 
the conducting phase. The thick lines represent the boundaries of the 
C component, dictated by equipotential contours near a symmetric saddle point 
of the potential; the dashed lines are the boundaries of C at percolation.
}
\end{figure}

{\em  The QH resistivity law.}
We focus on the transition from a $\nu=1$ liquid to the insulator \cite{IvsF}.
The $C$ component is an incompressible liquid at $\nu=1$, while $I$ consists 
of an electron solid of (lower) average filling fraction, $\nu_I$. 
$\nu_I$ depends on details such as the disorder 
potential, and hence is sample dependent \cite{Kukfn}. The  average filling 
fraction of the sample $\nu$ is
\beq
\nu=p +(1-p)\nu_I\; ,
\label{nueff}
\eeq
where $p$ is the area fraction  of the liquid. The percolation 
threshold in two dimensions is  at $p_c=0.5$.   
The  excess area  of the majority phase  near a saddle point is given by 
integrating between hyperbolas
 (see Fig. \ref{fig:fig1})
\be
\delta A={1\over 2 }\log(l_V/d) d^2
\ee
The total excess area fraction is thus related to the typical $d$ and $l_V$ by
\beq
p-p_c= \pm \gamma d^2,  ~~~~~~~\gamma=  N_{sp}\log(l_V/d)/(2A) 
\label{pvsd}
\eeq
where $A$ is total area of the sample and $N_{sp}$ is the number of
saddle points. Using (\ref{nueff}), 
(\ref{rland}), and(\ref{Tdtun}), we find that in both the 
insulator and liquid sides of the transition $\rho_{xx}(\nu)$ is given by the 
universal formula  Eq. (\ref{rlaw}), with $\nu_c=(1+\nu_i)/2$. The constants 
$\alpha$ and $\beta$ give the simplest interpolation formula
between the tunneling and activation regimes:   
\beq
\alpha= {8\gamma\over V''} (1-\nu_I)\quad
\beta={4l^2\gamma\over \pi} (1-\nu_I) \; .
\eeq

Note that the above analysis does not require extreme proximity to the 
percolation transition. The crucial assumption is
that the solid component of the quantum melt state is sufficiently insulating, 
such that the transport is dominated by a path that avoids it as much as 
possible.
The same assumption is necessary for observing a quantized Hall resistance, as 
discussed in \cite{SA}. This analysis therefore holds 
well beyond the  critical dynamical scaling regime.

{\em  Resistivity in Field Tuned Superconducting-Insulator transitions.}
The picture described above explains the remarkable
similarity of the empirical laws (\ref{rlaw}) and   (\ref{rlawB}). 
Both originate
from the Gaussian decay of transition rates at the saddle points.
For the superconducting side of the
field--tuned transition in amorphous MoGe \cite{YK}, 
we consider vortices crossing a narrow  superconducting channel of width  $d$.

The effects of internal interactions in the superconductor  
is provided by the first order line $\mu_{cr}(B)$. 
This allows us to relate the magnetic field to the width of the channel near 
the percolation field $B_{cr}$.
\be
{\partial \mu\over \partial B}\Bigg|_{B_{cr},\mu{cr}}( B_{cr}-B)=
{1\over 2} V'' d^2
\ee
which yields
\be
{\bar \alpha}={4\over  {\partial \mu\over \partial B_c}},\quad
{\bar\beta}={V''\over \pi^2 \rho_s {\partial \mu\over \partial B_c}}\; .
\ee 
One can obtain a semiquantitative estimate of $\rho_{xx}$ for the  amorphous 
MoGe data \cite{YK}
as follows.  At the critical field $B_{cr}$, 
there is a vortex lattice of spacing $\xi$ in the
superconductor. Consider a saddle point  channel which is pinched to zero 
width by two vortices at distance $\xi$. As the magnetic field is reduced 
their touching cores will separate by a distance 
$d= \sqrt{C\left(\phi_0^\ast /B -  \phi_0^\ast/B_{cr}\right)}$ 
where $C$ is a dimensionless constant of
order unity and  $\phi_0^\ast={h \over 2e c}$. 
Thus a superconducting channel of  width $d$ is formed.
Using the viscosity from Eq. 
(\ref{eta}), we obtain the zero temperature tunneling exponent, which yields
\be
\rho_{xx} \approx {h\over 4e^2} 
\exp\left[ C \pi/2 \left({\hbar/e^2\over \rho_n }\right) 
\left(B-B_{cr}\over B_{cr}\right)\right]
\label{rxx-theory}
\ee
We note that experiments of Ephron {\it et. al.} \cite{YK} have found very 
good agreement to (\ref{rxx-theory})
with $C\simeq 1.24$.

{\em Discussion:} Here we have used Boltzmann theory to  explain observed 
resistivity laws S-I and QH-I transitions.  
The absence of localization at zero temperature indicates a presence of 
strong dissipation. This allows us to neglect quantum interference effects at
long lengthscales, 
and justify the use of incoherent Boltzmann transport theory.
{\em However, the origin of this dissipation is not well understood. } 
One may expect that coupling to gapless Fermi liquid 
excitations would give rise to dissipation. 
But how could Fermi liquid excitations  be present in 
S-wave superconductors at zero temperature?  
``Normal'' electrons are recovered in mean field theory where
the BCS gap is destroyed by the magnetic field. However, if the 
{\em  local pair correlations} are present, one might prefer to  consider at 
the boundaries of the $S$ domains, a system of quantum disordered Cooper pairs 
subject to a penetrating field of $B\approx H_{c2}$. This
field puts approximately  one flux quantum per Cooper pair. A flux attachment 
transforms a Cooper pair into a composite fermion at $B=0$ 
\cite{CompositeFer}. A metallic state can thus be formed surrounding the $S$ 
domains which could be responsible for the resistive response at $T=0$. 

\acknowledgements
We thank  D. Chklovskii, N. Cooper, D. Gekhtman, B. I. Halperin, 
D. Shahar, S. Sondhi and D. C. Tsui for useful discussions.
This work was partly supported by the Technion -- Haifa University 
Collaborative
Research Foundation, the Fund for Promotion of Research at the Technion, 
grant no. 96--00294 from the United States--Israel Binational Science
Foundation (BSF), Jerusalem, Israel (E.S.), the Israel Science Foundation 
(A.A.) and NSF grant DMR 94-02131 (A.K.).

\begin{references}
\bibitem{FisherLee}
M. P. A. Fisher and D. H. Lee, Phys. Rev. B {\bf 39}, 2756 (1989);
 M. P. A. Fisher, Phys. Rev. Lett. {\bf 65}, 923 (1990);
D. H. Lee and M. P. A. Fisher, Int. J. of Mod. Phys. B, {\bf 5}, 2675 (1991).
\bibitem{SITrev}
A. F. Hebard, in {\it Strongly Correlated Electronic
Materials} (The Los Alamos Symposium 1993), edited by K. S. Bedell, Z. Wang,
D. E. Meltzer, A. V. Balatsky, and E. Abrahams, Addison Wesley (1994), p. 251;
G. T. Zimanyi, {\it ibid} p. 285;
Y. Liu and A. M. Goldman, Mod. Phys. Lett. B {\bf 8}, 277 (1994).
\bibitem{YK}
A. Yazdani and A. Kapitulnik, Phys. Rev.  Lett. {\bf 74}, 3037 (1995);
D. Ephron, A. Yazdani, A. Kapitulnik and M. R. Beasley, Phys. Rev.  Lett.
{\bf 76}, 1529 (1996).
\bibitem{Krav}
See also S. V. Kravchenko, W. E.
Mason, G. E. Bowker, J. E. Furneaux, V. M. Pudalov and M. D'Iorio, 
Phys. Rev. B {\bf 51}, 7038 (1995); S. V. Kravchenko, D. Simonian, M. P.
Sarachik, W. E. Mason and J. E. Furneaux, Phys. Rev. Lett. {\bf 77}, 4938
(1996).
\bibitem{KLZ}
S. Kivelson, D. H. Lee and S. C. Zhang, Phys. Rev. B {\bf 46}, 2223
(1992).
\bibitem{phas-tran}
For a review and extensive references, see A.~M.~M. Pruisken in 
{\it The Quantum Hall Effect,} Eds. R. E. Prange and S. M. Girvin
(Springer-Verlag, New York, 1986); S. Das Sarma in {\it Perspectives in the 
Quantum Hall Effect}, Eds. S. Das Sarma and A. Pinczuk (John Wiley and
Sons, 1997); the experimental situation is summarized in
S. L. Sondhi, S. M. Girvin, J. P. Carini and D. Shahar, 
Rev. Mod. Phys. {\bf 69}, 315 (1997).
\bibitem{Shahar1}
D. Shahar, D. C. Tsui, M. Shayegan, E. Shimshoni and S. L. Sondhi, 
Science {\bf 274}, 589 (1996); D. Shahar, D. C. Tsui, M. Shayegan,
J. E. Cunningham, E. Shimshoni an d S. L. Sondhi, Solid State Comm. {\bf 102},
817 (1997); D. Shahar, D. C. Tsui, M. Shayegan, E. Shimshoni
and S. L. Sondhi, Phys. Rev. Lett. {\bf 79}, 479 (1997).
\bibitem{Shahar2}
D. Shahar, M. Hilke, C. C. Li, D. C. Tsui, S. L. Sondhi and M. Razeghi,
preprint cond-mat/9706045.
\bibitem{nuprime}
In the case of a transition from a fractional QH state to the
insulator, it represents the filling fraction of composite fermions;
see E. Shimshoni, S. L. Sondhi and D. Shahar, Phys. Rev. B {\bf 55}, 13730
(1997).
\bibitem{Herb}
L. Zheng and H. A. Fertig, Phys. Rev. Lett. {\bf 73}, 878 (1994).
\bibitem{Chk}
D. B. Chklovskii, B. I. Shklovskii and L. I. Glazman, Phys. Rev. B {\bf 46}, 
4026 (1992); D. B. Chklovskii and P. A. Lee, Phys. Rev. B {\bf 48}, 
18060 (1993).
\bibitem{AA}  A. Aharony and A. Auerbach, Phys. Rev. Lett.
{\bf 70}, 1874 (1993).
\bibitem{MAA}
G. Murthy, D. Arovas and A. Auerbach,  Phys. Rev. B {\bf 55}, 3104 (1997).
\bibitem{LG}
P. Lam and S. M. Girvin,  Phys. Rev. B {\bf 30}, 473 (1984).
\bibitem{HLM}
B. I. Halperin, T. C. Lubensky and S.--K. Ma, Phys. Rev. Lett. 
{\bf 32}, 292 (1974).
\bibitem{PZ}
L. Pryadko and S.--C. Zhang,  Phys. Rev. Lett. {\bf 73}, 3282 (1994).
\bibitem{Kuk}
I. V. Kukushkin, V. I. Fal'ko, R. J. Haug, K. v. Klitzing and K. Eberl,
Phys. Rev. B {\bf 53}, 13260 (1996); see also 
D. Gekhtman, E. Cohen, A. Ron and L. N. Pfeiffer, Phys. Rev. B 
{\bf 54}, 10320 (1996).
\bibitem{Harvard}
Experimentally, that has been observed by
R. Willett et al., Phys. Rev. Lett. {\bf 59}, 1776 (1987); L. P.
Rokhinson, B. Su and V. J. Goldman, Solid State Comm. {\bf 96}, 309
(1995); Theoretical arguments were given by
I. M. Ruzin, N. R. Cooper and B. I. Halperin, Phys. Rev. B {\bf 53}, 
1558 (1996); N. R. Cooper, B. I. Halperin, C.--K. Hu and I. M. Ruzin,
preprint cond-mat/9608073
\bibitem{ruzin}
A. M. Dykhne and I. M. Ruzin, Phys. Rev. B {\bf 50}, 2369 (1994);
I. M. Ruzin and S. Feng, Phys. Rev. Lett. {\bf 74}, 154 (1995).
\bibitem{SA}
E. Shimshoni and A. Auerbach, Phys. Rev. B {\bf 58}, 9817 (1997).
\bibitem{Berg}
D. J. Bergman and D. Stroud, Solid State Physics {\bf 46}, 147 (1992).
\bibitem{KSCC}
in the context of QH transitions, critical aspects of such models were studied
by, e.g., J. Kucera and P. Streda, J. Phys. C {\bf 21}, 4357 (1988);
J. T. Chalker and P. D. Coddington, J. Phys. C {\bf 21}, 2665 (1988);
Y. Huo, R. E. Hetzel and R. N. Bhatt, Phys. Rev. Lett.
{\bf 70}, 481 (1993).
\bibitem{SITdual}
Duality symmetry is also observed in some S--I systems: in
Josephson arrays, by H. S. J. van der Zant, F. C. Fritschy, W. J. Elion,
L. J. Geerligs and J. E. Mooij, Phys. Rev. Lett. {\bf 69}, 2971 (1992); see
also \cite{Krav}.
\bibitem{cldual}
Classical  transport through  2D binary composites may also exhibit duality  
symmetry; see, e.g., J. B. Keller, J. Appl. Phys. {\bf 34}, 
911 (1963); J. Math.  Phys. {\bf 5}, 548 (1964). 
\bibitem{Assa}
A. Auerbach, preprint cond-mat/9707331.
\bibitem{RMP} 
G. Blatter {\it et al.}, Rev. of Mod. Phys. {\bf 66}, 1125 (1994).
\bibitem{BS} J. Bardeen and M. J. Stephen, Phys. Rev. {\bf 140}, A1197 (1965). 
\bibitem{IvsF}
Phenomenologically, it has been demonstrated \cite{Shahar1,nuprime}
that the experimental data at other QH transitions 
can be mapped to it using the correspondence rules of Ref. \cite{KLZ}.
\bibitem{Kukfn}
$\nu_I$ may also varry with $p$, however according to Ref. \cite{Kuk} $\nu$ 
is an approximately linear function of $p$ in a considerable  range around 
the transition, and hence we neglect this dependence.
\bibitem{CompositeFer} See, e.g., {\it Fractional Statistics and Anyon 
Superconductivity}, Ed. F. Wilczek (World Scientific, 1990), and references 
therein.


\end{references}
\end{document}